\documentclass[aps,pra,preprint]{revtex4}
\usepackage{graphicx}
\usepackage{dcolumn}
\usepackage{amsmath}
\usepackage{amssymb}
\begin{document}
\title{Comment on\\
{\bf Modifying the variational principle in the action integral
functional derivation of time-dependent density functional theory}\\ 
by Jochen Schirmer\\ } 
\author{Giovanni Vignale}
\affiliation{Department of Physics, University of Missouri-Columbia,
Columbia, Missouri 65211}
\date{\today}
\maketitle

In a paper recently published in Phys. Rev. A~\cite{Schirmer10} Schirmer has criticized an earlier work of  mine~\cite{Vignale08}, as well as the foundations of time-dependent density functional theory.  In Ref.[2], I showed that the so-called ``causality paradox"~\cite{vanLeeuwen}  -- i.e., the failure of the exchange-correlation potential derived from the Runge-Gross time-dependent variational principle~\cite{RG84} to satisfy causality requirements --  can be solved by a careful reformulation of that variational principle.  Fortunately,  the criticism presented  in Ref.[1] is based on elementary misunderstandings of the nature of functionals, gauge transformations, and the time-dependent variational principle.  In this Comment I wish to point out and clear these misunderstandings.

\begin{enumerate}
\item {\it Definition of the action functional --}
Ref.[1] claims that the action functional introduced by Runge-Gross~\cite{RG84} and adopted as the starting point of my work [2] is ill-defined because the wave function is defined up to an arbitrary time-dependent phase factor.  This is false, because multiplying  the wave function by a time-dependent phase factor $e^{-i\alpha [n] (t)}$, where $\alpha [n] (t)$ is an arbitrary functional of  the density, $n$, and  a function of time, $t$,  amounts to  adding to the Lagrangian the total time derivative $\frac{d\alpha [n](t)}{dt}$.  It is generally the case, both in classical and in quantum mechanics, that the Lagrangian is defined up to an arbitrary time derivative of a function of the coordinates and time~\cite{LL1}: it is well known that this ``gauge freedom" does not affect the {\it variation} of the action and therefore leaves the equations of motion unchanged.
In the present case,  the variational principle has the form (see Ref.~\cite{Vignale08}, Eq. (11))
\begin{equation}\label{VariationalPrinciple}
\delta A[n] =i\langle \psi_T[n]|\delta\psi_T[n]\rangle\,,
\end{equation}
where $A[n]$ is the action functional and $\psi_T[n]$ is the wave function, regarded as a functional of density and evaluated at the upper end $T$ of the time interval. Multiplying the wave function by the arbitrary phase factor
$e^{-i\alpha[n](t)}$ adds the same quantity to both sides of Eq.~(\ref{VariationalPrinciple}), i.e. we get,
\begin{equation}
\delta A[n] + \delta \alpha[n](T) =i\langle \psi_T[n]|\delta\psi_T[n]\rangle+\delta\alpha[n](T)\,.
\end{equation}
Therefore, the variational condition (\ref{VariationalPrinciple}) is completely unaffected by the arbitrary choice of phase.
The possibility of performing arbitrary gauge transformations is not a defect, but an expected and necessary feature of any correct theory.  

\item{\it Time-dependent variational principle -- }
Ref.[1] criticizes my statement of equivalence of the time-dependent variational principle to the Schr\"odinger equation.  It is argued that the vanishing of $\langle \delta \psi|i\partial_t - H|\psi\rangle$ does not necessarily imply the vanishing of $\left(i\partial_t - H\right)|\psi\rangle$.  In fact, this statement is wrong since the vanishing of $\langle \delta \psi|i\partial_t - H|\psi\rangle$  for arbitrary, unrestricted variations $\langle\delta \psi|$ does imply the vanishing of  $\left(i\partial_t - H\right)|\psi\rangle$.\cite{Langhoff}
Presumably, the author is thinking of  variations restricted to some parametrized subspace, but no such restriction was assumed implicitly or explicitly at the point where the statement was made in my paper, i.e., just after Eq. (6).

\item {\it Nature of functionals -- }
Ref.[1]  repeatedly asserts that the wave function should be a functional not only of the time-dependent density itself, but also of its derivatives with respect to time.  This assertion has its root in a lack of understanding of the difference between functionals and ordinary functions. A functional of the time derivative of the density is, by its very nature, a functional of the density, i.e. a mapping from the functional space of densities to the Hilbert space of wave functions.  The standard description of the wave function as a functional of the density~\cite{Parr-Yang} is therefore completely general and sufficient.  

\item{\it The ``loophole" -- }  In the last part of Ref.[1], Schirmer constructs a ``loophole" (his own word), whose only purpose is to cast doubts on my reformulation of the variational principle.  He suggests that the right hand side of my equation (11) (the second term on the left hand side of his Eq. (26)) might vanish for variations about certain particular exact densities.  This ``loophole" is either wrong or irrelevant since the fact that the right hand side of my Eq. (11) might vanish for certain densities in no way contradicts the validity of that equation.  What is worse, the subsequent discussion misleads the reader into thinking that the new formulation is hopelessly complicated.  In reality, I have shown in my paper that the new formulation does not introduce any additional complication, because the contribution of the new term on the right hand side of Eq. (11) cancels out when the functional derivatives of the functionals are calculated correctly.  
\end{enumerate}

In summary, Schirmer's critique of my paper is invalid, and my reformulation of the variational principle and the resolution of the causality paradox that follows from it stand in their pristine form.

\end{document}